\documentclass[12pt]{article}
\usepackage{graphicx}
\usepackage{amsmath}
\usepackage{hhline}
\usepackage{amssymb}
\usepackage{times}
\usepackage{cite}
\usepackage{array}
\usepackage{multirow}

\usepackage{color}
\definecolor{rltred}{rgb}{0.75,0,0}
\definecolor{rltgreen}{rgb}{0,0.5,0}
\definecolor{rltblue}{rgb}{0,0,0.75}

\newlength{\dinwidth}
\newlength{\dinmargin}
\begin{document}
\begin{titlepage}

\noindent
Date:        18 July 2016       \\
                
\vspace{2cm}

\begin{center}

\begin{LARGE}
{\bf VHEeP: A very high energy electron--proton collider}
\end{LARGE}

\vspace{2cm}

A. Caldwell$^1$ and M. Wing$^2$\\

\vspace{2cm}
$^1$Max Planck Institute for Physics, F\"{o}hringer Ring 6, 80805 Munich, Germany \\
$^2$Department of Physics and Astronomy, UCL, Gower Street, London, WC1E 6BT, UK \\

\end{center}

\vspace{2cm}

\begin{abstract}
  Based on current CERN infrastructure, an electron--proton collider is proposed at a centre-of-mass
  energy of about 9\,TeV.   A 7\,TeV LHC bunch is used as the proton driver to create a plasma wakefield
  which then accelerates electrons to 3\,TeV, these then colliding with the other 7\,TeV LHC proton beam.
  Although of very high energy, the collider has a modest 
  projected integrated luminosity of $10-100$\,pb$^{-1}$. For such a collider, with a centre-of-mass 
  energy 30 times greater than HERA, parton momentum fractions, $x$, down to about $10^{-8}$ are 
  accessible for photon virtualities, $Q^2$, of 1\,GeV$^2$.  The energy dependence of hadronic cross sections at high 
  energies, such as the the total photon--proton cross section, which has synergy with cosmic-ray physics, 
  can be measured and QCD and the structure of matter better understood in a region where the effects are 
  completely unknown.  
    Searches at high $Q^2$ for physics beyond the Standard Model will be 
  possible, in particular the significantly increased sensitivity to the production of leptoquarks.  
   These and other physics 
  highlights of a 
  very high energy electron--proton collider are outlined.
  \end{abstract}
  
%%%%  {\bf Relation saturation $\Leftrightarrow$ confinement ?}
\end{titlepage}

\section{Introduction}

The HERA electron--proton accelerator was the first and so far only lepton--hadron collider worldwide.  With 
its centre-of-mass energy of about 300\,GeV, HERA dramatically extended the kinematic 
reach~\cite{1506.06042} for the deep inelastic scattering process compared to fixed-target experiments.  
A broad range of physics processes were studied and new 
insights were gleaned from HERA which complemented the $p\bar{p}$ and $e^+e^-$  colliders, the Tevatron and 
LEP.  The LHeC project~\cite{lhec} is a proposed $ep$ collider with significantly higher energy and luminosity than 
HERA with a programme to investigate Higgs physics and QCD, to search for new physics, etc..  This will use 
significant parts of the LHC infrastructure at CERN with different configurations, such as $eA$, also possible.  
In this article, the possibility of having a very high energy electron--proton collider (VHEeP) is considered with an 
$ep$ centre-of-mass energy of about 9\,TeV, a factor of six higher than proposed for the LHeC and a factor of 30 
higher than HERA.

The VHEeP machine would strongly rely on the use of the LHC beams and the technique of plasma wakefield 
acceleration to accelerate electrons to 3\,TeV over relatively short distances.  Given such an acceleration scheme, 
the luminosity will be relatively modest with $10-100\,{\rm pb}^{-1}$ expected over the lifetime of the collider.  
With such an increase in centre-of-mass energy, the VHEeP collider will probe a new regime in deep inelastic 
scattering and QCD in general.  The kinematic regime accessible will be extended by three orders of magnitude 
compared to that measured at HERA.  This article puts forward the physics case for such a collider, highlighting 
some of the measurements most sensitive to unveiling new physics.  Complementary studies of high energy $ep$ 
colliders have been performed elsewhere~\cite{kaya1,kaya2,xia}, considering both the accelerator design~\cite{kaya1,xia} 
and physics potential~\cite{kaya2}.

The article is organised as follows.  In Section~\ref{sec:dis}, the kinematics and basic properties of deep inelastic 
scattering are defined.  In Section~\ref{sec:acc}, the scheme of plasma wakefield acceleration is briefly 
explained and a basic accelerator design for VHEeP is outlined, including justification of the centre-of-mass energy 
and estimate of the achievable luminosity.  The basic kinematics and properties of the final state at these new energies 
are discussed and also their effect on the choice of detector design are described in Section~\ref{sec:detector}.  In 
Sections~\ref{sec:qcd} and~\ref{sec:bsm}, the headline physics areas in QCD and beyond the Standard Model 
are outlined.  These include measuring the total photon--proton cross section, the deep inelastic scattering cross 
section at the lowest possible $x$ values and the search for leptoquarks.  In Section~\ref{sec:further}, the broad 
spectrum of other possible areas of study in high energy electron--proton collisions is briefly given.  In 
Section~\ref{sec:summary}, the physics case is 
summarised and an outlook for the VHEeP project given.

\section{Deep inelastic scattering}
\label{sec:dis}

Deep inelastic scattering~\cite{devenish:cooper-sarkar} can be classified as either neutral current (NC) or charged 
current (CC), depending on whether 
the exchanged boson is a photon or $Z^0$ boson (NC) or a $W^\pm$ boson (CC).  The NC reaction, which dominates the 
cross section for the processes considered here, in electron--proton 
collisions can be written as $e^- (k) + p(p) \to e^-(k^\prime) + X(p^\prime)$ with the four vectors given in the brackets 
and $X$ referring to a hadronic final state.  In the case of CC, the final state electron is replaced by a neutrino.  The 
four-momentum of the exchanged boson, $q$, is given by $q=k-k^\prime$.  The events can be described by the following 
quantities: the squared centre-of-mass energy, $s$, is given by

\begin{equation}
s = (k+p)^2\,;
\end{equation}
the virtuality, $Q^2$, of the exchanged boson is given by

\begin{equation}
Q^2 = -q^2\,;
\end{equation}
the Bjorken-$x$ variable, interpreted as the fraction of proton's momentum carried by the struck quark in the proton's 
infinite-momentum frame, is given by

\begin{equation}
x = \frac{Q^2}{2 p \cdot q}\,;
\end{equation}
the inelasticity, $y$, interpreted in the proton rest frame as the fraction of energy transferred from the lepton to the proton, is given by

\begin{equation}
y = \frac{p \cdot q}{p \cdot k}\,.
\end{equation}
The above variables are related, $Q^2 = s\,x\,y$, where particle masses can be ignored at the very high energies considered here.  
Another quantity of importance is the exchanged-boson--proton centre-of-mass energy, $W$, which is given by

\begin{equation}
W^2 = (q+p)^2\,.
\end{equation}

The neutral current cross section for $e^- p$ scattering, $\frac{d^2\sigma^{e^- p}_{\rm NC}}{dx\,dQ^2}$, can be written in terms 
of these variables and the structure functions of the proton as

\begin{equation}
\frac{d^2\sigma^{e^- p}_{\rm NC}}{dx\,dQ^2} = \frac{2\,\pi\,\alpha^2}{x\,Q^4}\, \left( Y_+ F_2 - Y_- xF_3 - y^2 F_{\rm L} \right)\,,
\end{equation}
where $\alpha$ is the fine structure constant and $Y_\pm = 1 \pm (1-y)^2$.  The structure function $F_2$ is sensitive to the quark and 
antiquark distributions in the proton and dominates at low $Q^2$; the structure function $xF_3$ is sensitive to the difference in 
the quark and antiquark distributions in the proton and arises due to the interference of the photon- and $Z^0$-exchange contributions 
at $Q^2$ values around the mass of the $W$ and $Z$ bosons.  The longitudinal structure function $F_{\rm L}$ is sensitive to the 
gluon distribution in the proton and becomes important at high values of $y$.

The double-differential cross section can also be written in terms of the photon--proton cross section, $\sigma^{\gamma p}$, and the 
photon flux, $\phi$.  The equivalent photon approximation~\cite{epa} relates these as

\begin{equation}
\frac{d^2 \sigma^{e^- p}}{dy \, dQ^2} = \phi(y, Q^2) \sigma^{\gamma p}(y, Q^2)\,,
\end{equation}
where all quantities depend on both $Q^2$ and $y$.

The event kinematics are calculated using the initial beam energies and the kinematics of the scattered electron and hadronic final state.  
Of particular importance are the scattered electron energy, $E_e^\prime$, and polar angle, $\theta_e$, and the hadronic angle, 
$\gamma_{\rm had}$, where both angles are measured with respect to the proton beam direction.

\section{VHEeP accelerator complex}
\label{sec:acc}

Tajima and Dawson first proposed that plasmas can sustain very large electric fields capable of accelerating bunches of 
particles~\cite{prl:43:267}.  Using a laser pulse or electron bunch to drive the plasma ``wakefield", accelerating 
gradients of up to, respectively, 100\,GV\,m$^{-1}$~\cite{pwa-laser} and 50\,GV\,m$^{-1}$~\cite{pwa-e} have 
been measured.  This acceleration concept can also make use of bunches of protons~\cite{pdpwa}, given  
high energy proton bunches are available and hence the possibility to have the acceleration performed in 
one stage.  Simulation has shown that the plasma wakefield created by the LHC proton bunches can 
accelerate a trailing bunch of electrons to 6\,TeV in 10\,km~\cite{pp:18:103101}.  The concept of proton-driven 
plasma wakefield acceleration will be tested by the AWAKE collaboration at CERN which aims to demonstrate 
the scheme for the first time~\cite{awake}.  The initial aims of the AWAKE experiment are to demonstrate GeV 
acceleration of electrons within 10\,m of plasma~\cite{awake}.  Following this, the AWAKE collaboration proposes to 
accelerate bunches of electrons to 10\,GeV in about 10\,m of plasma~\cite{awake-erik-ipac}.

\begin{figure}[h]
\begin{center}
\includegraphics[trim={7.5cm 3cm 11.cm 4cm},clip,width=0.5\textwidth]{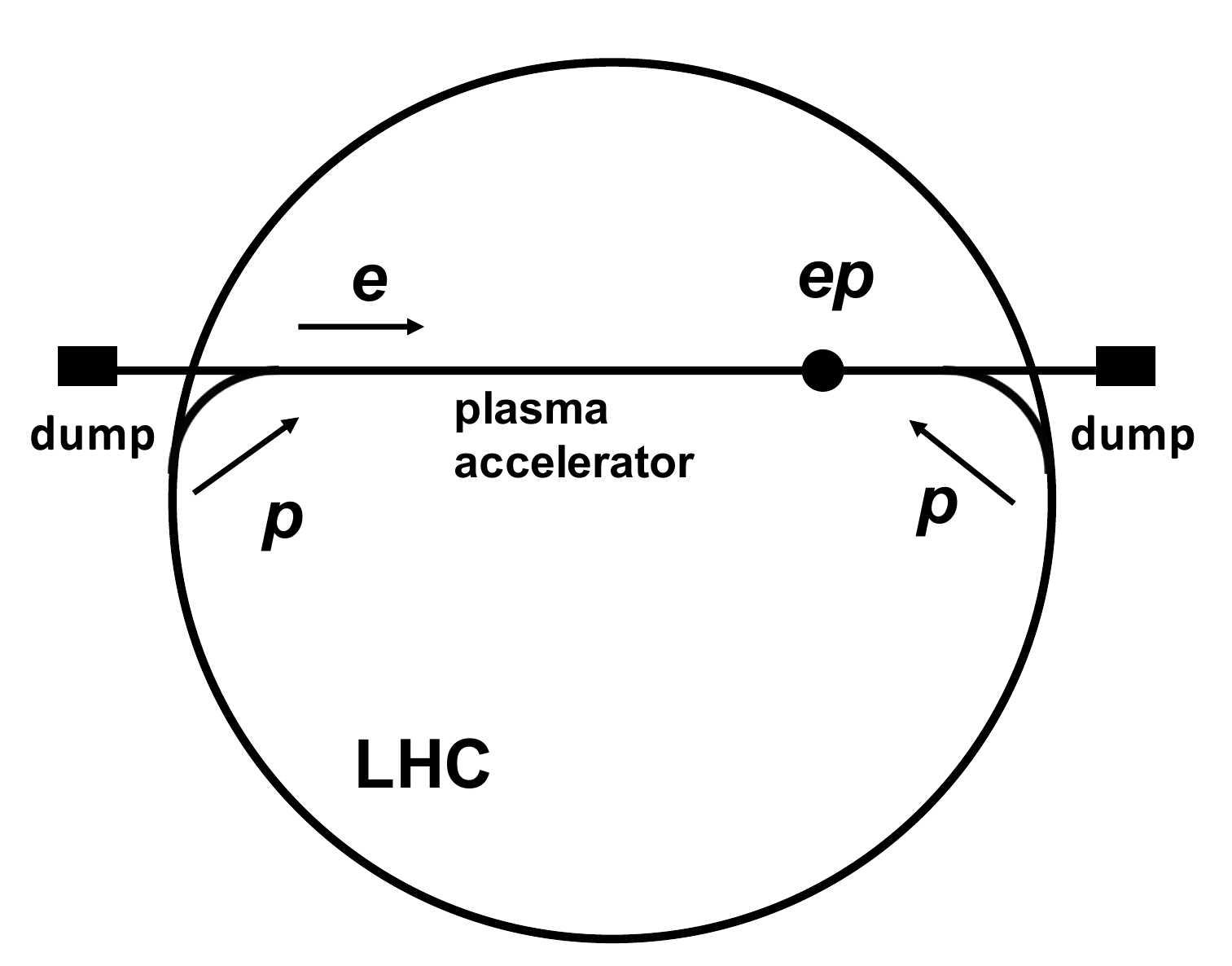}
\caption{Simple schematic of the VHEeP accelerator complex, showing the LHC ring.  Protons from the LHC are 
extracted into the VHEeP plasma accelerator and used to accelerate bunches of electrons.  Proton bunches 
rotating in the other direction in the LHC are extracted into the VHEeP tunnel and collided with electrons.  
Both proton and electron dumps could be used for fixed-target (beam-dump) experiments.
}
\label{fig:acc}
\end{center}
\end{figure}

The very high energy electron--proton collider (VHEeP) is based on current LHC infrastructure and a new tunnel to 
house the plasma accelerator.  The facility uses one 
of the LHC proton beams to generate wakefields and accelerate a trailing electron bunch which then collides with 
the other proton beam.  This is shown in a simple schematic in Fig.~\ref{fig:acc} in which the electron beam is 
chosen to have an energy of 3\,TeV, achieved in a plasma accelerator of $\lesssim$\,4\,km, and the proton beams have an 
energy of 7\,TeV.  Separation of the drive proton beam and witness electron beam will be needed to avoid $pp$ collisions; 
as well as the temporal difference, the beams will need to be separated transversely by $O$(mm).  Further study of this 
important issue is needed.

Although the energy of VHEeP will be very high, obtaining high luminosities will be, as with all plasma wakefield acceleration 
schemes, a challenge.  In these initial studies, an integrated luminosity over the lifetime of VHEeP of $10-100$\,pb$^{-1}$ is 
considered~\cite{dis-proc}, based on the expected capabilities of the LHC and pre-accelerators.  The lower limit will be sufficient for 
measurements at low $x$ where the cross section is expected to rise with decreasing $x$.  A higher integrated luminosity 
will aid the search for physics beyond the Standard Model, typically at high $Q^2$.

\section{VHEeP kinematics and basic detector design}
\label{sec:detector}

In order to investigate the kinematic distributions of events at these high energies, a small sample of deep inelastic scattering 
events was generated using the {\sc Ariadne} Monte Carlo programme~\cite{cpc:71:15,zfp:c65:285} with requirements 
$Q^2 > 1$\,GeV$^2$, $W^2 > 5$\,GeV$^2$ and $x > 10^{-7}$ for a luminosity of about 0.01\,pb$^{-1}$.  The 
CTEQ2L~\cite{pl:b304:159} set of proton parton distribution functions was used as this gave a reasonable and continuous 
distribution at low $x$.  The cut on $x$ was required because of technical difficulties generating events down to $x = 10^{-8}$, 
the kinematic limit for $Q^2 > 1$\,GeV$^2$, as the parton density functions in CTEQ2L are not valid at these very low values. 

\begin{figure}[h]
\begin{center}
\includegraphics[width=0.8\textwidth]{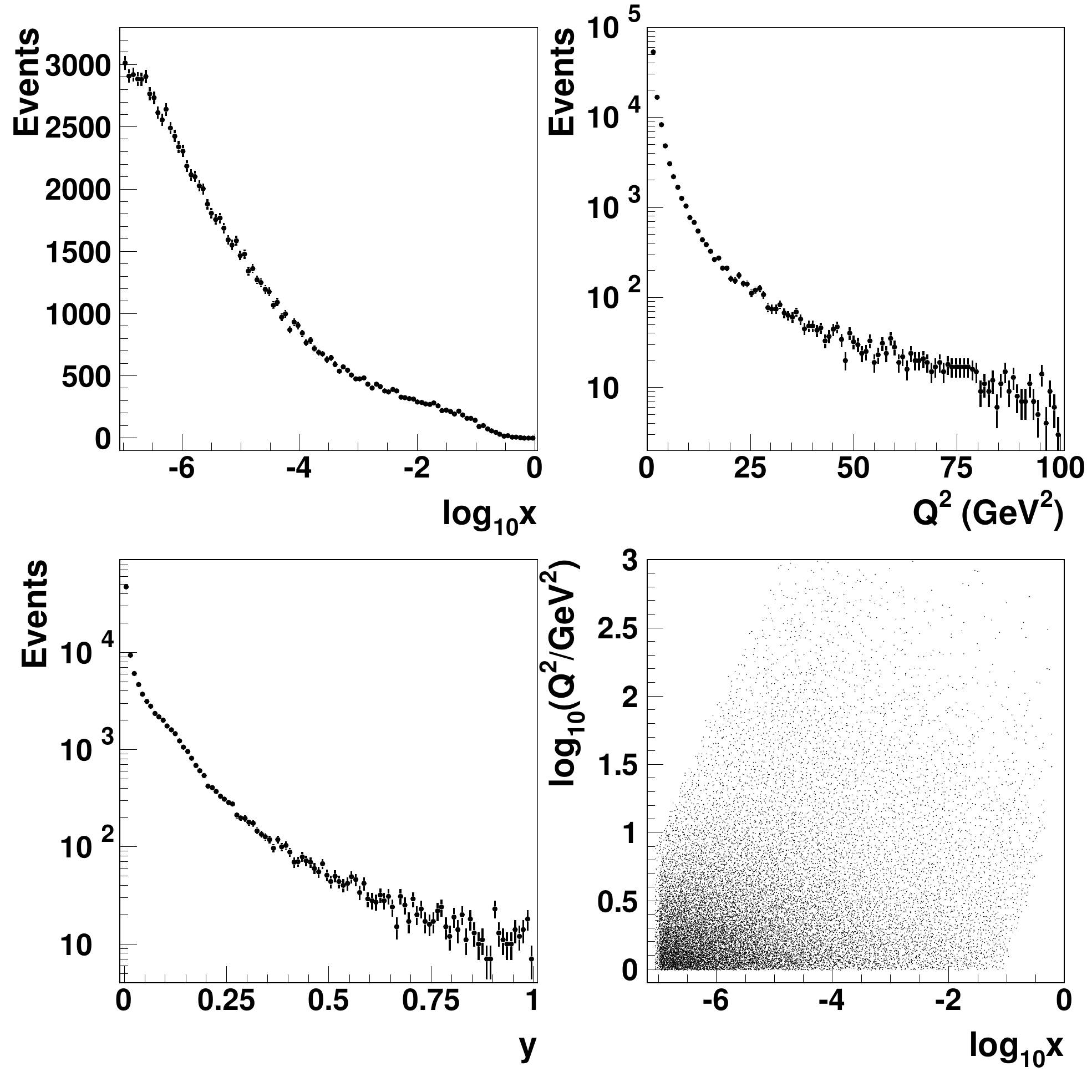}
\caption{Distributions of $\log_{10} x$, $Q^2$, $y$ and $\log_{10} Q^2$ versus $\log_{10} x$ generated using the 
{\sc Ariadne} Monte Carlo programme for $Q^2 > 1$\,GeV$^2$, $W^2 > 5$\,GeV$^2$ and $x > 10^{-7}$ for a luminosity of  
0.01\,pb$^{-1}$.  
}
\label{fig:dis-kinematics}
\end{center}
\end{figure}

The basic kinematic distributions of deep inelastic scattering events are shown in Fig.~\ref{fig:dis-kinematics}, focusing on the 
low-$x$ and low-$Q^2$ region.  In this region, a high luminosity will not be needed due to the high cross section, with e.g.\ 10s of 
million of events in the region $10^{-7} < x < 10^{-6}$ for $Q^2 > 1$\,GeV$^2$ and an integrated luminosity of 10\,pb$^{-1}$.  
It should be noted that 
the lowest value of $Q^2$ measured at HERA was $Q^2 = 0.045$\,GeV$^2$, which at VHEeP corresponds to a minimum $x$ 
value of $5 \times 10^{-10}$.  At this $Q^2$, a significantly larger number of events is expected.

\begin{figure}[h]
\begin{center}
\includegraphics[width=0.8\textwidth]{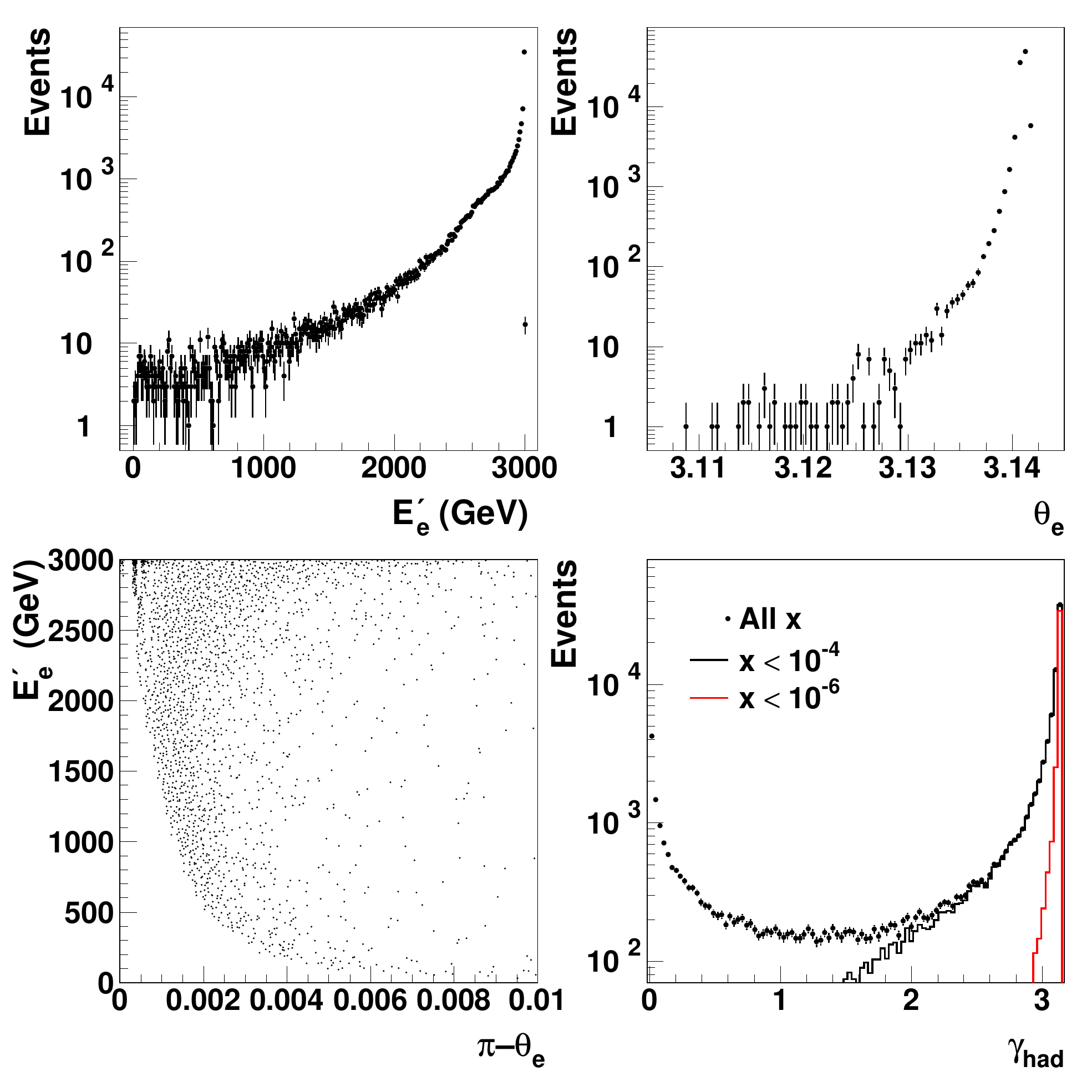}
\caption{Distributions of scattered electron energy, $E_e^\prime$, scattered electron angle, $\theta_e$, $E_e^\prime$ versus 
$\pi - \theta_e$, and 
hadronic angle, $\gamma_{\rm had}$, generated using the {\sc Ariadne} Monte Carlo programme for 
$Q^2 > 1$\,GeV$^2$, $W^2 > 5$\,GeV$^2$ and $x > 10^{-7}$ for a luminosity of 0.01\,pb$^{-1}$.  The hadronic 
angle is also shown for different requirements on $x$.}
\label{fig:kinematics}
\end{center}
\end{figure}

Kinematic distributions of the electron and hadron final state are shown in Fig.~\ref{fig:kinematics}.  The scattered electron 
energy, $E_e^\prime$, is strongly peaked at the initial electron beam energy of 3\,TeV, but with a tail down to a few GeV.  
The angle of the scattered electron, $\theta_e$, measured with respect to the proton beam, is shown to strongly peak at 
180$^\circ$, with the higher the energy, the stronger the peak, as shown in the correlation between $E_e^\prime$ and 
$\pi - \theta_e$.  The angle of produced hadrons, $\gamma_{\rm had}$, is 
distributed over 0$^\circ$ to 180$^\circ$, but with a peak at 0$^\circ$ and even stronger peak at 180$^\circ$.  The events at 
low angles are due to events at high $x$, whereas the events with hadrons at high angles are dominated by events at low 
$x$.

\begin{figure}[htp]
\includegraphics[trim={0.cm 5.2cm 0cm 4.2cm},clip,width=\textwidth]{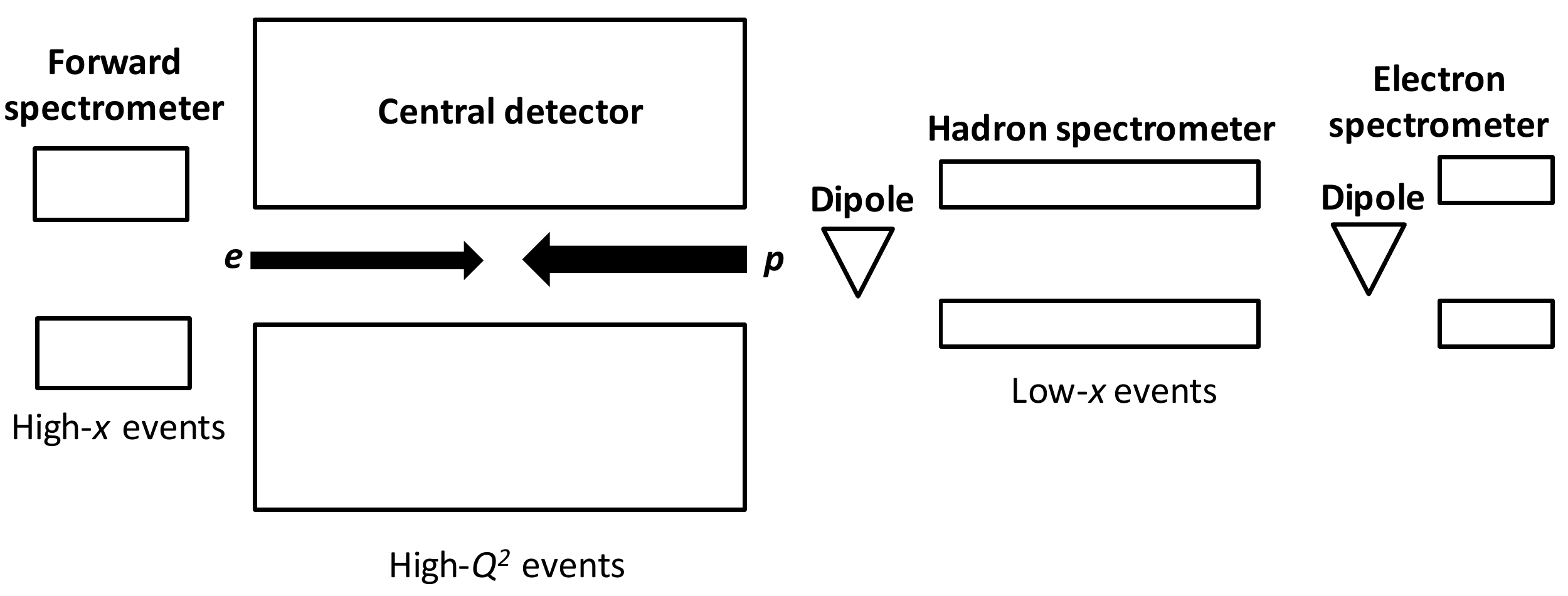}
\caption{Simple schematic of the detector needed for VHEeP, with a central detector, and extended spectrometer 
arms for, in particular, high-$x$ and low-$x$ events.  This overall design could be replicated for more than one $ep$ 
collision point. 
}
\label{fig:detector}
\end{figure}

The distributions in Fig.~\ref{fig:kinematics} have consequences for the detector design with a central detector needed as well as 
instrumentation close to the beamline to measure both electrons and hadrons.  A simple schematic of the detector needed for 
VHEeP is shown in Fig.~\ref{fig:detector}.  A central detector, which is expected to be similar to other colliding-beam experiments, 
will be needed to reconstruct the hadronic final state and, in particular, events at high $Q^2$.  Additionally, long spectrometer arms 
will be required in the electron direction to measure the hadronic final state at low $x$ and the scattered electron.  A spectrometer 
in the direction of the proton beam will be required to measure the hadronic final state at high $x$. The systems will possibly 
consist of dipole magnets 
to extract the particles from the beamline and low-angle detector systems.  Clearly, an understanding of the radiation created by 
the beams as well as possible radiation from the plasma acceleration will need detailed study.

\section{QCD physics at VHEeP}
\label{sec:qcd}

Electron--proton collisions at $\sqrt{s} \sim 9$\,TeV give access to a completely new kinematic regime for deep inelastic 
scattering with, in particular, a reach in $x$ a factor of about 1\,000 lower than at HERA and, depending on the luminosity, 
a similar increase in the reach at high $Q^2$.  The energy dependence of hadronic cross sections, such as the total 
photon--proton cross section, are poorly understood.  Predictions calculated from first principles are often not available 
and so phenomenological models are used to describe the dependence.  Being able to measure the energy dependence, 
particularly with the long lever arm presented by VHEeP, will deepen our understanding of QCD and the structure of matter.
In this section, some highlight physics measurements which particularly benefit from this extended kinematic regime are 
discussed.

\subsection{Total photon--proton cross section}
\label{sec:total}

\begin{figure}[h]
\includegraphics[width=\textwidth]{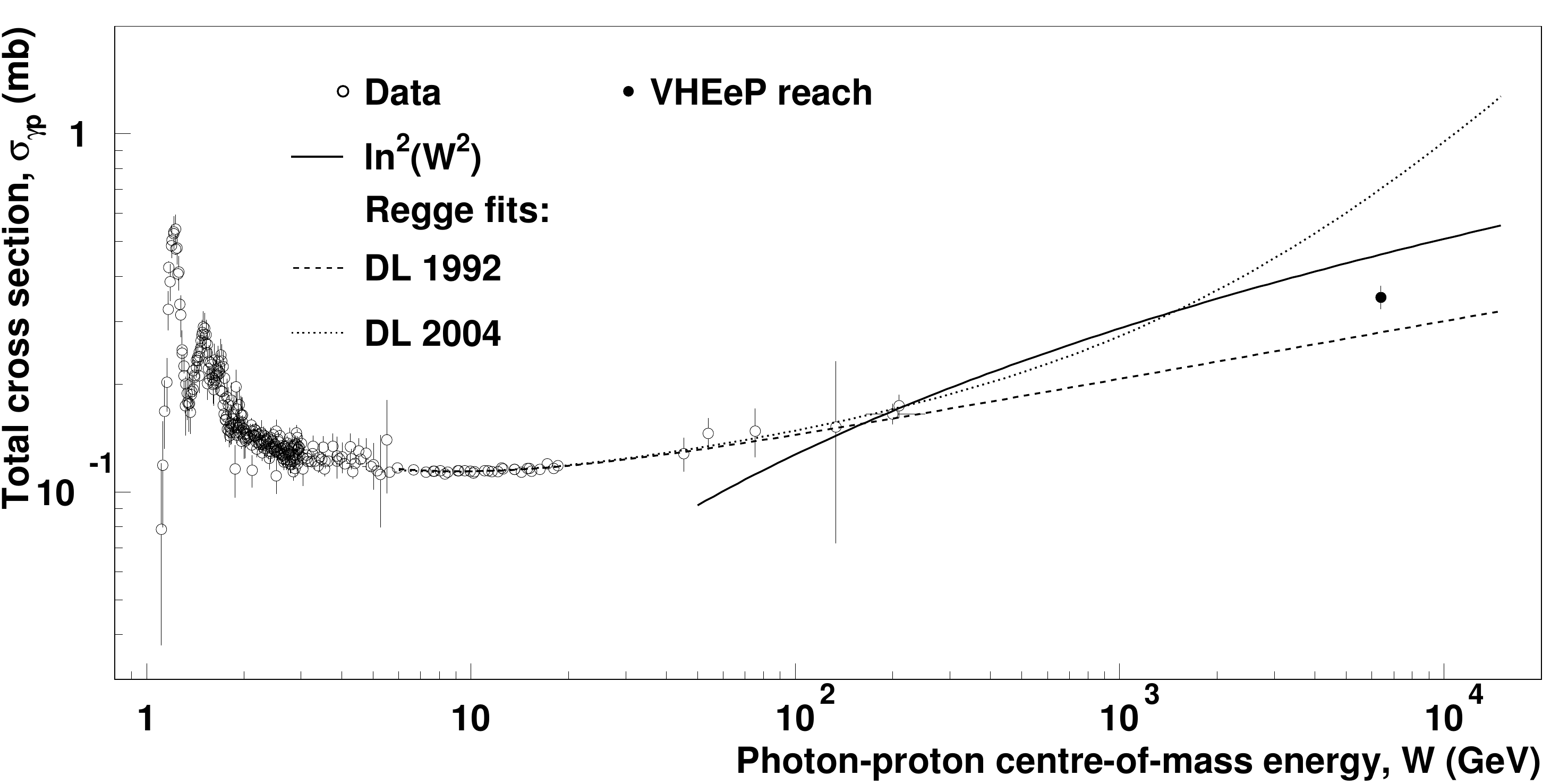}
\caption{Total $\gamma p$ cross section versus photon--proton centre-of-mass energy, $W$, shown for data compared to 
various models.  The data is taken from the PDG~\cite{pdg}, with references to the original papers given therein. The 
VHEeP data point is shown at the same $W$ value relative to $\sqrt{s}$ as the HERA results.  The 
VHEeP cross section is assumed to be double the ZEUS value and the same uncertainties are assumed.  The ZEUS 
measurement is at $\sqrt{s} = $209\,GeV and used a luminosity of 49\,nb$^{-1}$. The Regge fits shown are those of 
Donnachie--Landshoff (1992~\cite{pl:b296:227} and 2004~\cite{Donnachie:2004pi}). The $\ln^2(W^2)$ form is based  
on~\cite{pr:123:1053}.}
\label{fig:gammap}
\end{figure}

Measurements of the total $\gamma p$ cross section are shown in Fig.~\ref{fig:gammap} compared to phenomenological 
models.  A projection for a measurement at the expected maximum $W$ at VHEeP is also shown; further measurements 
of similar precision at lower $W$ are also expected.  The many cross sections at low energies can be fitted in 
Regge phenomenology in which the dominant contributions arise from Reggeon exchange, which falls with increasing 
centre-of-mass energy, and Pomeron exchange, which rises.  As an example, two such fits from Donnachie and 
Landshoff~\cite{pl:b296:227,Donnachie:2004pi} 
are shown in the figure. The 1992 fit predates HERA data and is made over the range $6 < W < 20$\,GeV.  It has a single 
Pomeron term and a single Reggeon term. The 2004 fit includes HERA data (photoproduction, as well as DIS data for 
$Q^2 < 45$\,GeV$^2$) and allows an additional second Pomeron term.  Up to the highest HERA energies, both fits give 
good descriptions of the data.  At higher energies, the second Pomeron term starts to become dominant and the 
cross-section predictions differ significantly.  Also shown is an extrapolation based on the Froissart bound~\cite{pr:123:1053}, 
$\sigma_{\gamma p} \propto \ln^2(s)$.  Assuming measurements from VHEeP up to 
$W \sim 6$\,TeV, even with very low luminosities ($\mathcal{L} = 49$\,nb$^{-1}$, as used in the ZEUS 
measurement~\cite{np:b627:3}), the data will be able to strongly constrain the energy dependence of the total cross section 
and hence provide a clearer picture of QCD.  It should be noted  
that the multi-TeV energies and hence large lever arm attainable at VHEeP are necessary to do this.

A photon--proton collision of $W = 6$\,TeV, corresponding to photon and proton energies of, respectively, 1.3\,TeV and 7\,TeV, 
is equivalent to a 20\,PeV photon on a fixed target.  This extends significantly 
into the region of ultra high energy cosmic rays.  Therefore VHEeP data could be used to constrain cosmic-ray air-shower 
simulations and so will be of benefit to understanding the nature of cosmic rays at the highest energies (for example, 
see~\cite{pr:d92:114011}).  

\subsection{Vector meson production}
\label{sec:vm}

\begin{figure}[htp]
\begin{center}
\includegraphics[trim={0.52cm 0cm 0cm 5cm},clip,width=0.45\textwidth]{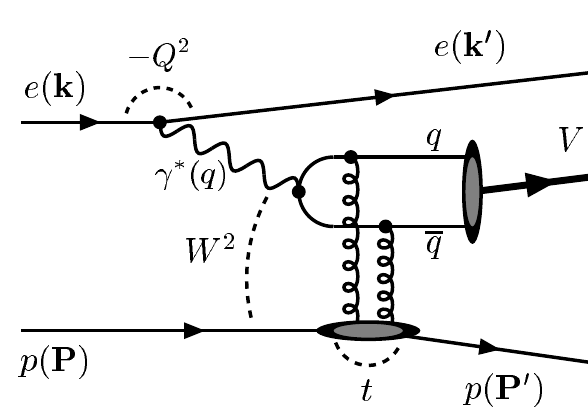}
\caption{Feynman representation of vector meson production in $ep$ collisions.}
\label{fig:vm-feyn}
\end{center}
\end{figure}

Vector meson production is dependent on the partonic distributions in the proton and given the need for two gluons from the 
proton to create a vector meson, see Fig.~\ref{fig:vm-feyn}, is particularly sensitive to saturation of the parton densities or 
other effects.  For high vector meson masses, QCD calculations are expected to be more reliable and so measurements 
of $J/\psi$ production can constrain the gluon density in a complementary way to fits to inclusive cross section in deep 
inelastic scattering (see~\cite{epj:c24:345,epj:c73:2466} and references therein).  

\begin{figure}[htp]
\begin{center}
\includegraphics[width=0.8\textwidth]{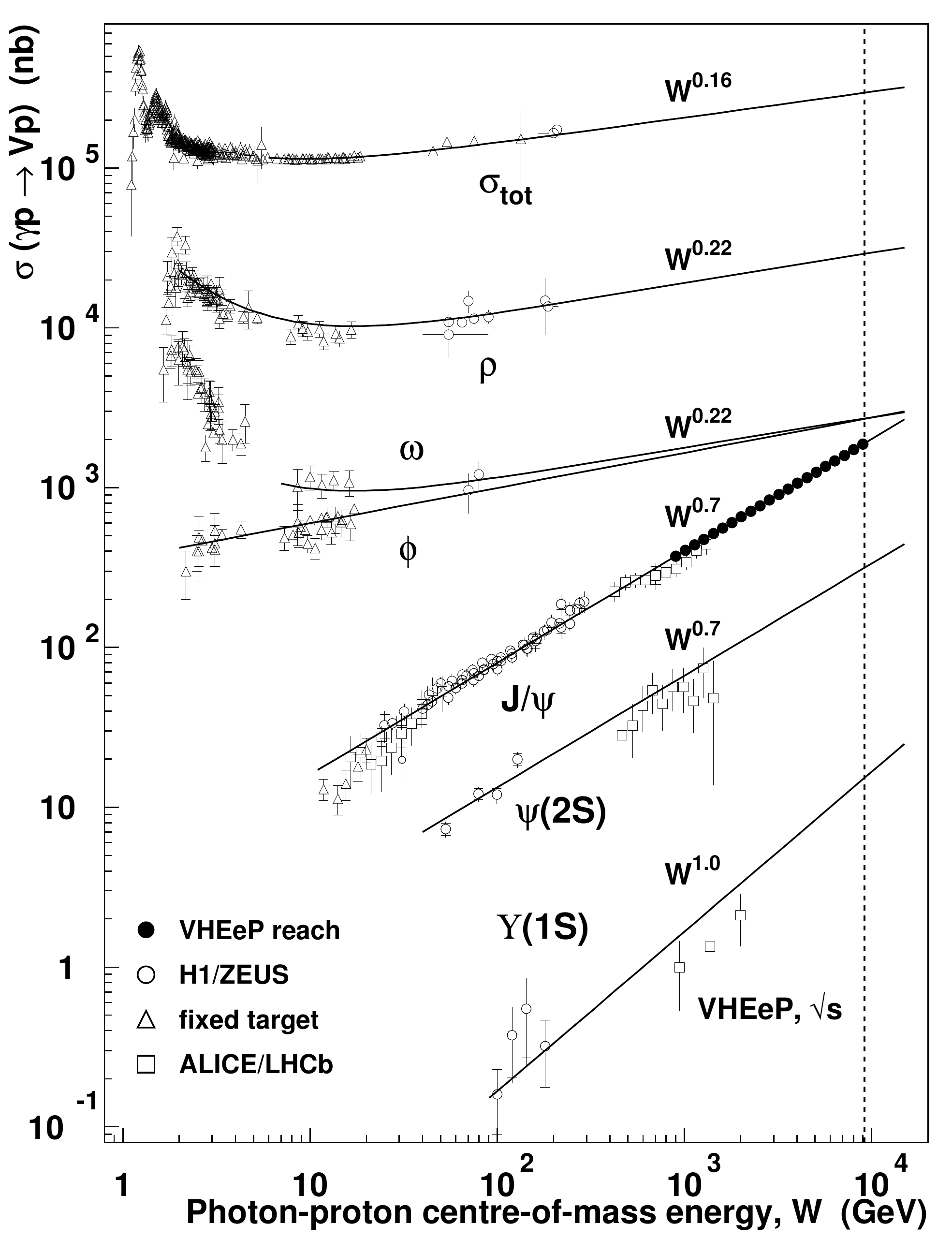}
\caption{The total $\gamma p$ cross section compared to the cross sections for exclusive vector meson production.  The 
data in each of the reactions is shown along with a function depending on a power of $W$.  The ALICE and LHCb data are 
extracted from measurements of vector meson production in $p-Pb$ and $pp$ collisions, respectively~\cite{lhc:vm}.  The reach of VHEeP is 
shown with simulated points for $J/\psi$ production up to the centre-of-mass energy, indicated as the vertical dashed line.}
\label{fig:vm}
\end{center}
\end{figure}

Vector meson production has been measured extensively at HERA and fixed-target experiments~\cite{hera-ft:vm} and the 
cross sections are shown in 
Fig.~\ref{fig:vm} as a function of the photon--proton centre-of-mass energy, $W$.  Photoproduction cross sections extracted from 
LHC data~\cite{lhc:vm} are also shown, which extend up to the TeV scale.  The cross sections all rise with increasing $W$ or equivalently 
decreasing $x$.  These cross section dependences, as well as the total cross section from 
Fig.~\ref{fig:gammap} which is also shown, can be parametrised in terms of a power of $W$, as also shown in 
Fig.~\ref{fig:vm}.  The simple power-law behaviour of $W$ with particle mass describe well the data from fixed-target, 
HERA and LHC experiments.  However, such functions lead to non-sensical results at VHEeP energies, also indicated, in which 
e.g.\ $\sigma_{J/\psi}$ approaches $\sigma_\omega, \sigma_\phi$ and will at some point be larger.  Clearly, the cross sections 
must take on another form and start to level out, be it through saturation or some other mechanism.  As shown, data from VHEeP 
will be able to determine this behaviour.

\subsection{Physics at low Bjorken {\boldmath $x$}}
\label{sec:lowx}

As discussed above, the energy dependence of the total photoproduction cross section at high energies is of great interest, 
both on fundamental grounds and for understanding cosmic-ray events in the atmosphere.  The energy dependence of 
scattering cross sections for virtual photons on protons is also of fundamental interest, and its study at different virtuality 
is expected to bring insight into the processes leading to the observed universal behaviour of cross sections at high 
energies.

In deep inelastic scattering of electrons on protons at HERA, the strong increase in the proton structure function $F_2$ with 
decreasing $x$ for fixed, large, $Q^2$ is usually interpreted as an increasing density of partons in the proton, providing more 
scattering targets for the electron.  This interpretation relies on choosing a particular reference frame to view the scattering -- 
the Bjorken frame.  In the frame where the proton is at rest, it is the state of the photon or weak boson that differs with varying 
kinematic parameters.  For the bulk of the electron--proton interactions, the scattering process involves a photon, and we can 
speak of different states of the photon scattering on a fixed proton target.  What is seen is that the photon--proton cross section 
rises quickly with $W$ for fixed $Q^2$~\cite{ref:sigmagp}.  In the proton rest 
frame, we interpret this as follows: as the energy of the photon increases, time dilation allows shorter lived fluctuations of the 
photon to become active in the scattering process, thereby increasing the scattering cross section.  

\begin{figure}[htp]
\includegraphics[width=\textwidth]{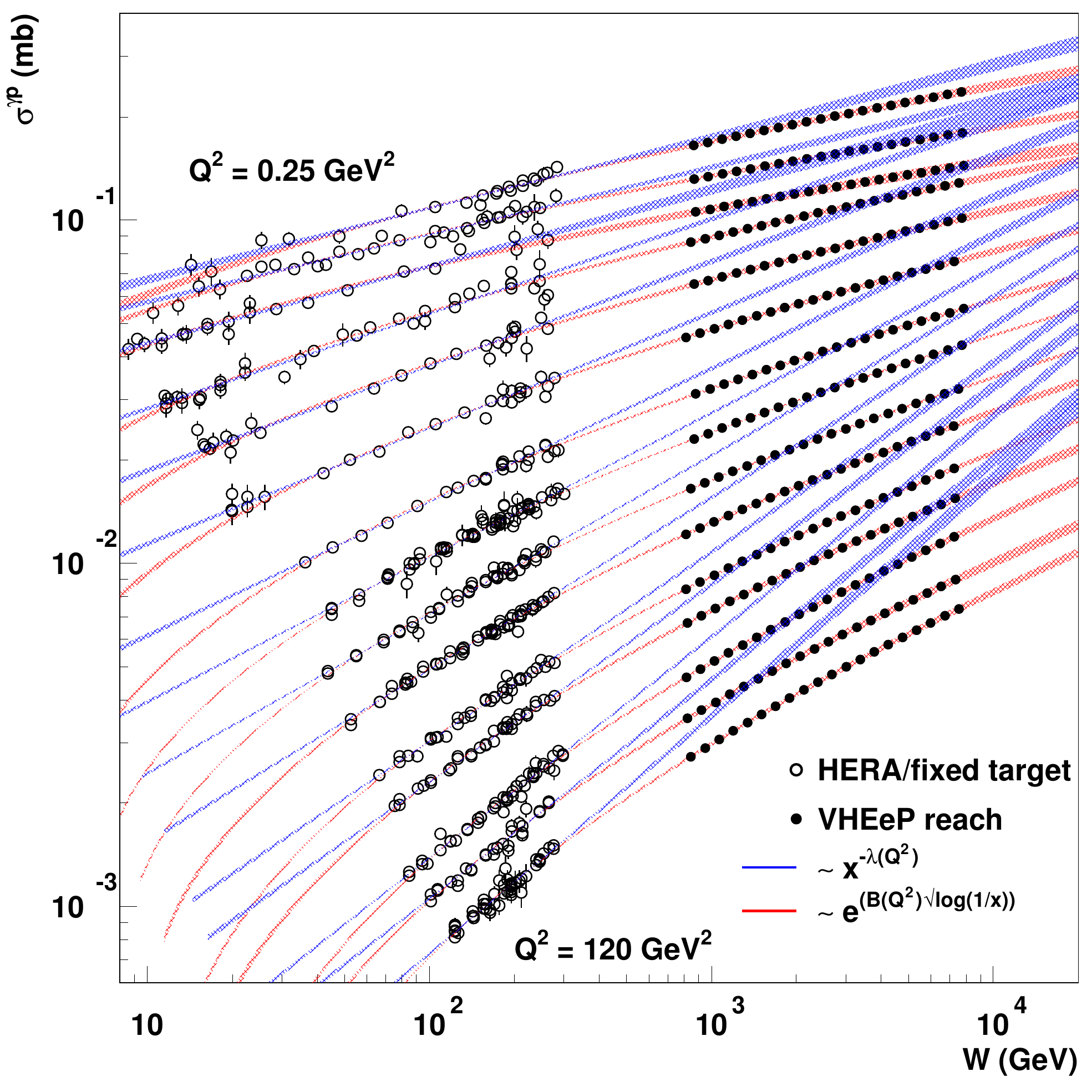}
\caption{Measurements (open points) of $\sigma^{\gamma p}$ versus $W$ for $0.25 < Q^2 < 120$\,GeV$^2$ from HERA and 
fixed-target experiments.  
The blue lines show fits to the data, performed separately for each $Q^2$ value, of the form given in Eq.~\ref{eq:smallx}.  
The red lines show fits of the form given in Eq.~\ref{eq:das}.  The reach of VHEeP is shown as projected data points (closed points).  
The points are placed on the red curve.  The uncertainties are assumed to be of order 1\%, given the increased cross section 
expected and similar systematics to those at HERA and are not visible as error bars on this plot.}
\label{fig:lowx}
\end{figure}

Figure~\ref{fig:lowx} shows the results of extrapolation of fits to the energy dependence of the photon--proton cross section 
for different virtualities as given in the caption~\cite{ref:caldwell}, for two different assumptions on the energy behaviour.  In one 
instance (blue curves), the energy dependence is assumed to follow a simple behaviour at small values of $x$:

\begin{equation}
\sigma^{\gamma p} \propto x^{-\lambda(Q^2)}\,,
\label{eq:smallx}
\end{equation}
while in the second instance a form inspired by double asymptotic scaling~\cite{ref:BallForte} was used 

\begin{equation}
\sigma^{\gamma p} \propto e^{B(Q^2)\cdot\sqrt{\log{1/x}}} \; .
\label{eq:das}
\end{equation}
The simple behaviour is what has been used in most fits to HERA data~\cite{1506.06042} to date, and it deviates strongly 
from the expectations of double asymptotic scaling in the VHEeP kinematic range.  It is found that this simple behaviour 
cannot continue to ever smaller values of $x$ as this would result in large-$Q^2$ cross sections becoming larger than 
small-$Q^2$ cross sections.  A change of the energy dependence is therefore expected to become visible in the VHEeP 
kinematic range.  This should yield exciting and unique information on the fundamental underlying physics at the heart 
of the high energy dependence of hadronic cross sections.

\section{Physics beyond the Standard Model}
\label{sec:bsm}

We now switch to a discussion of prospects for the discovery of new physics beyond the Standard Model (BSM).  The key 
features of VHEeP are the very large centre-of-mass energy and the unique combination of electron and proton scattering.  
We therefore focus on new physics where these features are essential.  Two topics were chosen for this initial study: the 
search for quark substructure, and the search for production of leptoquarks.  This is by no means an exhaustive list of topics, 
but is intended to show the capabilities of VHEeP for BSM physics.

\subsection{Quark substructure}
\label{sec:rq}

In order to look for quark substructure, an effective finite radius of the quark can be assigned.  The Standard Model prediction, 
$\frac{d\sigma^{\rm SM}}{dQ^2}$, is modified using a semi-classical form-factor approach~\cite{form-factor}:

\begin{equation}
\frac{d\sigma}{dQ^2} = \frac{d\sigma^{\rm SM}}{dQ^2} \left( 1 - \frac{R_e^2}{6}Q^2 \right)^2 \left( 1 - \frac{R_q^2}{6}Q^2 \right)^2  \, ,
\label{eq:rq}
\end{equation}
where $R_e^2$ and $R_q^2$ are the mean-square radii of the electron and quark radius, respectively.  As has been done in a recent 
analysis of HERA data~\cite{paper:rq}, the radius of the electron is assumed to be zero, i.e.\ the electron is point-like.

The existence of a finite quark radius primarily changes the $Q^2$ distribution; in a more complete analysis, the full 
$x-Q^2$ dependence of the cross section must be analysed to separate the effects of a finite quark radius from effects 
due to uncertain parton density distributions.  For this first study, it was assumed that the parton densities were known 
with negligible uncertainty, and only the modification of the $Q^2$ dependence was considered.  The effect of the finite 
quark radius grows with $Q^2$, as seen in Eq.~\ref{eq:rq} but this is counterbalanced by the rapidly falling cross section 
with $Q^2$.  Deep inelastic scattering events were generated with the {\sc Ariadne} Monte Carlo programme, 
corresponding to a luminosity of $100$\,pb$^{-1}$.  The event distribution as a function of $Q^2$  for the range 
$10^5 < Q^2 < 10^7$\,GeV$^2$ was analysed and an upper limit on $R_q$ was determined by reweighting the known 
cross section using Eq.~\ref{eq:rq} and performing a fit to the simulated data using the BAT package~\cite{ref:BAT}.  The 
limits correspond to $68$\,\% credibility upper limits where a flat prior was taken for $0\leq R_q<5 \cdot 10^{-19}$\,m.

The extracted limit on $R_q$ is $R_q\leq 1 \times 10^{-19}$\,m, which can be compared with the $95$\,\% Confidence 
Level limit extracted from HERA data~\cite{paper:rq}, $R_q<4 \times 10^{-19}$~m.  The limit extracted from the HERA data 
was the result of a much fuller analysis; it is expected that the limits from VHEeP would become considerably stronger if 
lower $Q^2$ data were included in the analysis, but this would require a much more complete analysis using detailed 
information on systematic uncertainties.  The limit would also improve by about a factor $3$ for a factor $10$ increase in luminosity.

\subsection{Leptoquark production}
\label{sec:lq}

\begin{figure}[htp]
\begin{center}
\includegraphics[trim={0cm 11cm 10cm 0cm},clip,width=0.5\textwidth]{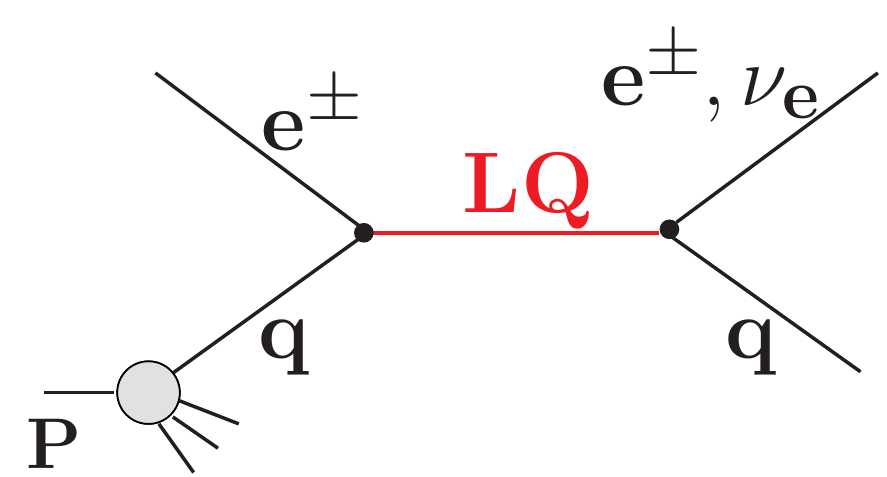}
\caption{Feynman representation of $s$-channel production of a leptoquark in $ep$ collisions.}
\label{fig:lq-feyn}
\end{center}
\end{figure}

Electron--proton collisions are particularly sensitive to leptoquark production as the leptoquark is produced resonantly in the 
$s$-channel.  This is shown pictorially in Fig.~\ref{fig:lq-feyn}, where an electron and quark fuse, with a coupling $\lambda$.  
The leptoquark subsequently decays to a lepton--quark system, again with a coupling $\lambda$, and this effect can be 
searched for by reconstructing the invariant mass of the final states or looking for a resonant deviation from the Standard 
Model in the $x$ distribution which is related to the mass of the leptoquark.

\begin{figure}[htp]
\includegraphics[width=0.49\textwidth]{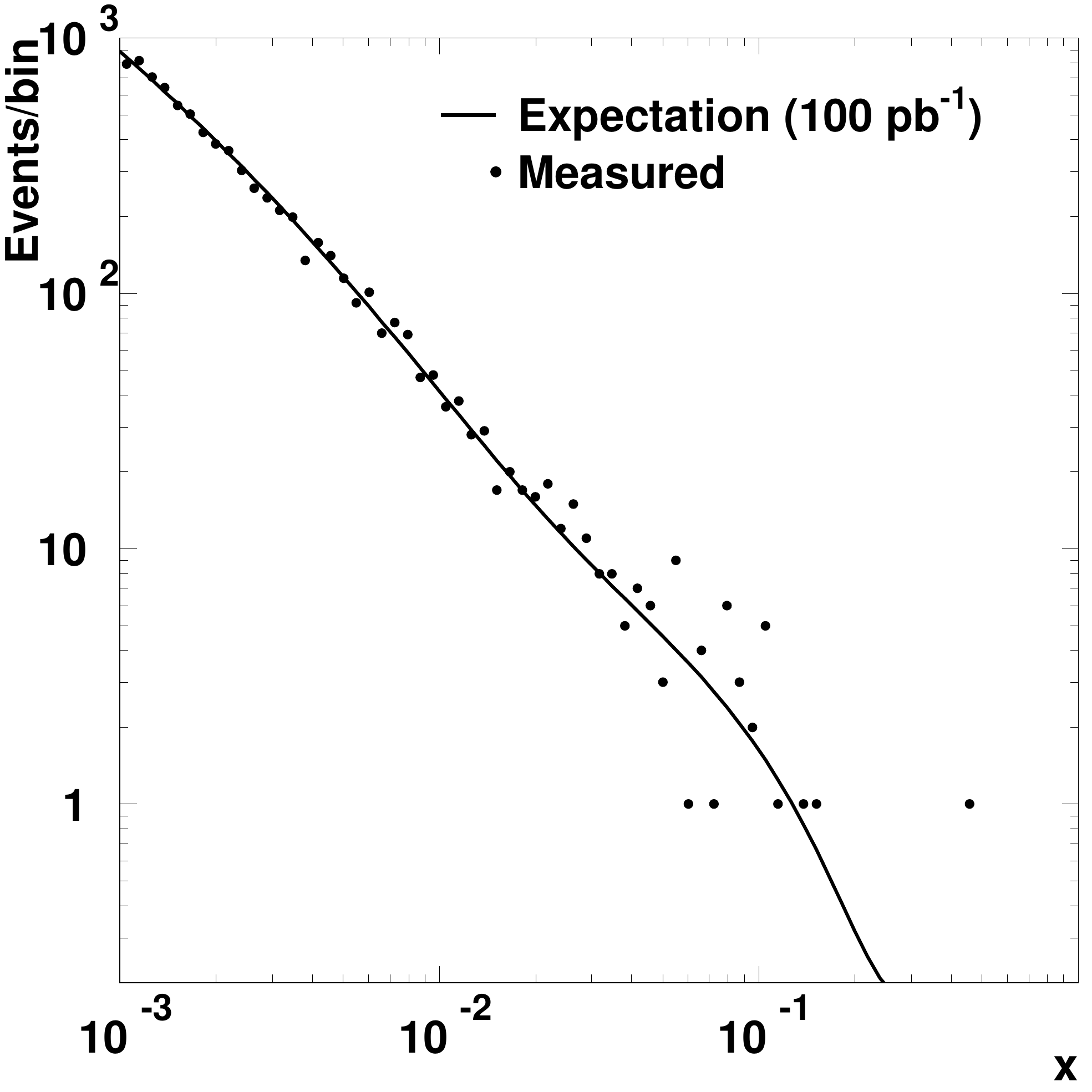}
\put(-7,70){\makebox(0,0)[tl]{(a)}}
\put(70,70){\makebox(0,0)[tl]{(b)}}
\includegraphics[width=0.49\textwidth]{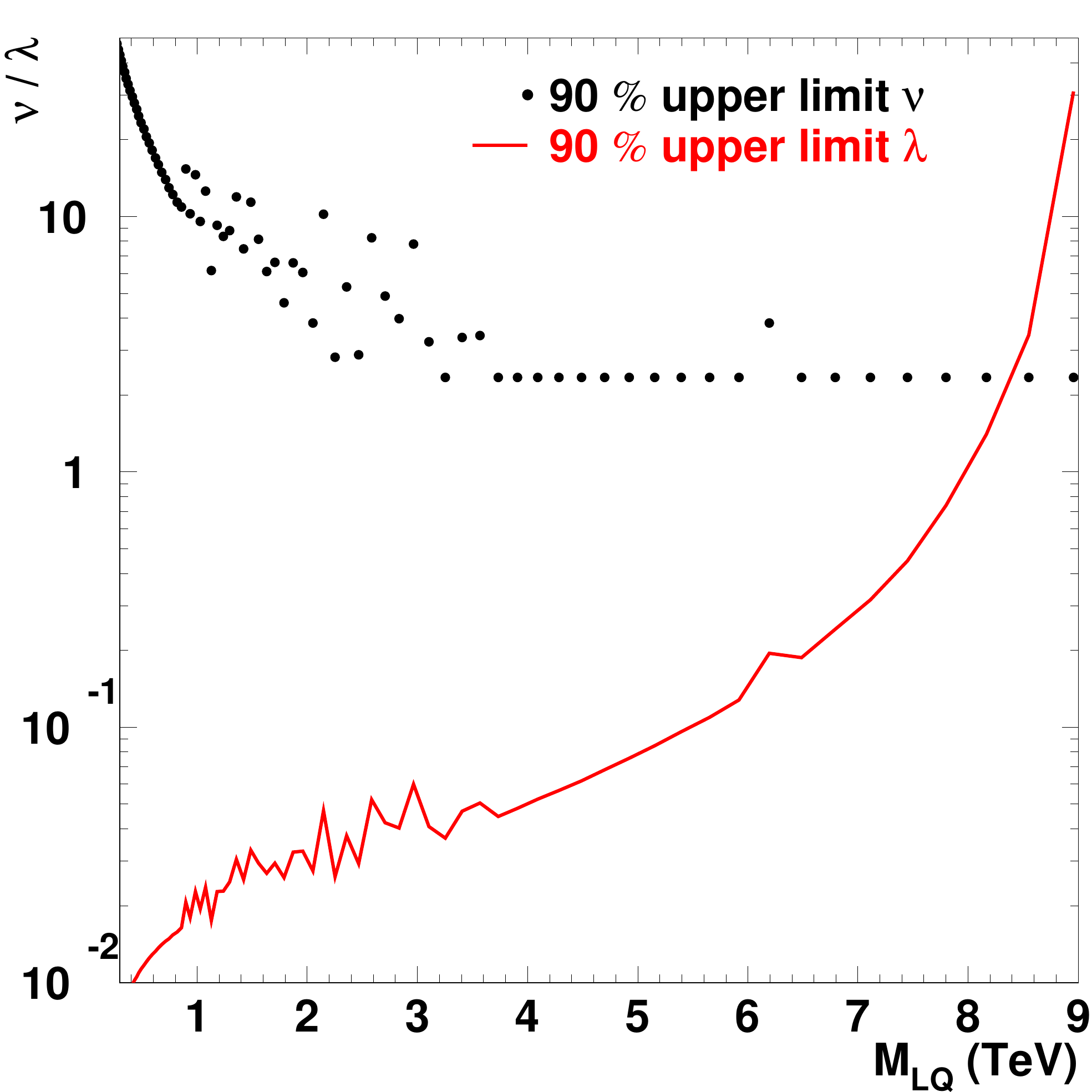}
\caption{(a) Simulated deep inelastic scattering data with a luminosity of 100\,pb$^{-1}$ (points) and the expectation, both 
generated with the {\sc Ariadne} Monte Carlo programme with $Q^2 > 10\,000$\,GeV$^2$ and $y>0.1$.  (b) Upper limits 
on the number of events, $\nu$, and leptoquark coupling parameter, $\lambda$, versus mass of the leptoquark, 
$M_{\rm LQ}$. }
\label{fig:lq}
\end{figure}

In this analysis, deep inelastic scattering events were generated with the {\sc Ariadne} Monte Carlo programme and the $x$ 
distribution plotted.  This is the same sample as used to extract the limit on $R_q$, corresponding to a luminosity of 
100\,pb$^{-1}$, with events up to about $x \sim 0.5$, see Fig.~\ref{fig:lq}(a).  Cuts $Q^2 > 10\,000$\,GeV$^2$ and $y>0.1$ 
were applied to enhance the possible signal over background.  A much larger independent sample was generated, again 
using {\sc Ariadne}, and used as the Standard Model prediction, also shown in Fig.~\ref{fig:lq}(a).  The 90\% probability 
upper limit on the number of signal events, $\nu$, above the Standard Model prediction was then extracted based on this 
pseudo-data sample and is shown as a function of the leptoquark mass in Fig.~\ref{fig:lq}(b).

In order to extract a signal or limit on leptoquark production, the Standard Model prediction is convoluted with the prediction 
for leptoquark production according to the Buchm\"{u}ller--R\"{u}ckl--Wyler (BRW) model~\cite{BRW}.  The Born-level cross 
section for resonant $s$-channel leptoquark production in the narrow-width approximation (NWA), $\sigma^{\rm NWA}$, is

\begin{equation}
\sigma^{\rm NWA} = (J+1) \frac{\pi}{4\,s}\lambda^2 \, q(x_0,M^2_{\rm LQ})
\end{equation}
where $q(x_0,M^2_{\rm LQ})$ is the initial-state quark (or antiquark) parton-density function in the proton for a Bjorken-$x$ value of 
$x_0=M^2_{\rm LQ}/s$, where $M_{\rm LQ}$ is the mass of the leptoquark, and $J$ is the spin of the leptoquark.  Given the 
limit on $\nu$, a limit on the coupling $\lambda$ as a function of the mass of scalar leptoquarks was extracted and is 
shown in Fig.~\ref{fig:lq}.

These results show that VHEeP has sensitivity up to the kinematic limit of 9\,TeV.  The HERA 
limits are $\lambda=0.01$ just below the kinematic limit of 0.3\,TeV, rising rapidly to $\lambda=1$ at 
about 1\,TeV~\cite{pl:b704:388,pr:d86:012005}.  Limits from the LHC experiments are also $\lambda=1$ at 
about $1-2$\,TeV for $pp$ collisions at $\sqrt{s} = 8$\,TeV for $\mathcal{L} \sim 20$\,fb$^{-1}$~\cite{epj:c76:5,pr:d93:032005}.  
Given the increased centre-of-mass energy and higher luminosities the sensitivity to leptoquark production at the LHC will 
extend to $2-3$\,TeV.  Hence VHEeP has a sensitivity to leptoquark production significantly beyond the HERA limits 
and LHC limits, both measured and expected.

\section{Further physics areas}
\label{sec:further}

As well as the areas discussed in the previous sections, many other areas, in particular in QCD, will be open to 
investigation at VHEeP.  Standard tests of QCD, performed at HERA and other colliders, will be possible in this 
new kinematic regime, such as measurements of the strong coupling constant, jet and heavy flavour production 
and properties of the hadronic final state.  The structure of the proton and photon can be further investigated and 
fits of the parton densities to the data performed.  The results will be related to the measurements above, sensitive 
to effects such as saturation, but more conventional determinations of the inclusive deep inelastic scattering cross 
section can be made 
in a region less sensitive to the more exotic QCD effects.  In particular, given the possibility to change the beam 
energy and to also vary this as widely as possible, the longitudinal structure function can be well measured, 
which was done with only limited precision at HERA given the relatively small lever arm in centre-of-mass 
energy~\cite{epj:c74:2814,pr:d90:072002}.  Given the possibility to run with other heavy ions as well as protons 
in the LHC, VHEeP will also be able to investigate the properties of electron--ion, $eA$, scattering.  Finally, diffraction, 
an area of QCD re-invigorated at HERA, and particularly sensitive to low-$x$ dynamics can be studied.

\section{Summary and outlook}
\label{sec:summary}

The concept of proton-driven plasma wakefield acceleration will be tested in the next few years in the AWAKE 
experiment at CERN, with first results expected already this year.  Recent simulations of the scheme indicate 
that acceleration of electrons with $7$\,TeV LHC proton bunches could be used to bring electrons to $3$\,TeV 
with an average gradient of about $1$\,GeV/m.   This opens up tremendous opportunities for considering very 
high energy colliders; in this paper, we consider the possibility of colliding $3$\,TeV electrons accelerated in 
this way with $7$\,TeV protons from the LHC. This will allow an extension of the kinematic reach of the HERA 
electron--proton collider by three orders of magnitude, albeit with moderate luminosities.

It is wise to decouple achieving very high energies in a particle collider from the requirement for very high 
luminosities.  While cross sections for standard $s$-channel physics lead to extremely high luminosity 
requirements for TeV and beyond energy scales, there are physics questions which can be probed with lower 
luminosities.  Since achieving high luminosities with realistic power consumption could be technologically 
much more difficult than achieving high energies, it is important that the physics case for a high energy but 
low luminosity collider be investigated.  We consider just such an option in this paper -- a very high energy 
electron--proton collider, VHEeP.  We have investigated a number of physics topics that could be addressed 
by VHEeP and found that indeed very fundamental particle physics questions could be addressed by such a 
collider.  These range from clarifying the underlying physics leading to the energy dependence of cross 
sections at very high energies, including unraveling the mechanisms for a saturation of the cross section 
growth, to opening new windows for physics beyond the reach of the LHC, such as leptoquark production 
with masses beyond $3$\,TeV. The studies presented in this paper are just a small first step in understanding 
the physics possibilities of such a collider, and we heartily encourage our colleagues to consider other topics.  
While we believe that the collider parameters given in this paper are achievable, there is a long road ahead 
to realise such a machine.  New ideas could well arise along the way that would allow for higher luminosities 
than we have assumed here. Clearly, serious studies of the accelerator and detector will be needed.  We 
believe these are warranted by the exciting physics that a collider such as VHEeP would provide.

\section*{Acknowledgements}
L. L\"{o}nnblad is 
gratefully acknowledged for discussions using the {\sc Ariadne} Monte Carlo programme in this new kinematic 
regime.  E. Shaposhnikova is also gratefully acknowledged for discussions on the LHC beam parameters.  R. McNulty 
is gratefully acknowledged for assistance with the LHCb data on vector meson production.
M. Wing acknowledges the support of STFC, DESY and the Alexander von Humboldt Stiftung.

\end{document}